\documentclass[11pt,a4paper]{article}
\pdfoutput=1

\usepackage[utf8]{inputenc}
\usepackage[english]{babel}
\usepackage{chicago}

\usepackage[round]{natbib}
\usepackage[margin=1in]{geometry}

\usepackage{graphicx}
\usepackage{amsmath, amsfonts, amssymb, verbatim, url}

\usepackage{amsthm}
\theoremstyle{plain}% default
\newtheorem{proposition}{Proposition}
\theoremstyle{definition}
\newtheorem{definition}{Definition}
\theoremstyle{remark}
\newtheorem{remark}{Remark}

\renewcommand{\vec}[1]{\boldsymbol{#1}}
\renewcommand{\hat}[1]{\widehat{#1}}
\newcommand{\x}{\vec{x}}
\newcommand{\T}{^{\top}}
\newcommand{\inv}{^{-1}}
\newcommand{\mub}{\vec{\mu}}
\newcommand{\betab}{\vec{\beta}}
\newcommand{\thetab}{\vec{\theta}}
\newcommand{\Sigmab}{\vec{\Sigma}}
\newcommand{\Sigmabar}{\bar{\Sigmab}}
\newcommand{\I}{\vec{I}}
\newcommand{\D}{\vec{D}}
\newcommand{\A}{\vec{A}}
\newcommand{\C}{\vec{C}}
\renewcommand{\a}{\vec{a}}
\newcommand{\ones}{\vec{1}}
\newcommand{\Exp}{\mathrm{E}}
\newcommand{\Var}{\mathrm{Var}}
\newcommand{\Cov}{\mathrm{Cov}}
\newcommand{\diag}{\mathrm{diag}}
\newcommand{\Real}{\mathbb{R}}
\newcommand{\ind}{\bot\hspace*{-6pt}\bot} 
\newcommand{\V}{\vec{V}}
\renewcommand{\v}{\vec{v}}

\renewcommand{\L}{\vec{L}}
\newcommand{\X}{\vec{X}}
\newcommand{\Z}{\vec{Z}}
\newcommand{\M}{\vec{M}}
\usepackage{mathrsfs}
\newcommand{\Set}{\mathscr{S}}

\newcommand{\Space}{\mathcal{S}}
\newcommand{\Model}{\mathcal{M}}
\newcommand{\BICdiff}{BIC_{\mathrm{diff}}}
\newcommand{\BICclust}{BIC_{\mathrm{clust}}}
\newcommand{\BICnotclust}{BIC_{\mathrm{not\,clust}}}
\newcommand{\BICreg}{BIC_{\mathrm{reg}}}

\usepackage[usenames]{color}
\definecolor{DarkBlue}{rgb}{0.1,0.1,0.5}
\definecolor{DarkBlue2}{rgb}{0.42,0.44,0.58}
\definecolor{Black}{rgb}{0.0,0.0,0.0}
\definecolor{Red}{rgb}{0.9,0.0,0.1}
\definecolor{DarkRed}{rgb}{0.6,0.00,0.08}
\definecolor{DarkGreen}{rgb}{0.00,0.6,0.08}
\begin{document}

\title{Dimension reduction for model-based clustering}
\author{Luca Scrucca \\ Universit\`a degli Studi di Perugia}
\date{\today}

\maketitle

\begin{abstract}
We introduce a dimension reduction method for visualizing the clustering structure obtained from a finite mixture of Gaussian densities. Information on the dimension reduction subspace is obtained from the variation on group means and, depending on the estimated mixture model, on the variation on group covariances.
The proposed method aims at reducing the dimensionality by identifying a set of linear combinations, ordered by importance as quantified by the associated eigenvalues, of the original features which capture most of the cluster structure contained in the data. 
Observations may then be projected onto such a reduced subspace, thus providing summary plots which help to visualize the clustering structure. 
These plots can be particularly appealing in the case of high-dimensional data and noisy structure.
The new constructed variables capture most of the clustering information available in the data, and they can be further reduced to improve clustering performance. We illustrate the approach on both simulated and real data sets.

\end{abstract}

%%%%%%%%%%%%%%%%%%%%%%%%%%%%%%%%%%%%%%%%%%%%%%%%%%%%%%%%%%%%%%%%%%%%%%%

\section{Introduction}
\label{intro}
Model-based clustering assumes that the observed data are generated from a mixture of $G$ components, each representing the probability distribution for a different group or cluster \citep{McLachlan:Peel:2000}. 
The general form of a finite mixture model is 
$f(\x) = \sum_{g=1}^G \pi_g f_g(\x|\thetab_g)$, 
where the $\pi_g$'s are the mixing probabilities such that $\pi_g \ge 0$ and $\sum\pi_g=1$, $f_g(\cdot)$ and $\thetab_g$ are, respectively, the density and the parameters of the $g$-th component ($g=1,\ldots,G$).
With continuous data, we often take the density for each mixture component to be the multivariate Gaussian $\phi(\x|\mub_g,\Sigmab_g)$ with parameters $\thetab_g = (\mub_g,\Sigmab_g)$. Thus, clusters are ellipsoidal, centered at the means $\mub_g$, and with other geometric features, such as volume, shape and orientation, determined by $\Sigmab_g$.

Parsimonious parametrization of covariance matrices can be adopted  through eigenvalue decomposition in the form $\Sigmab_g = \lambda_g \D_g \A_g \D\T_g$, where $\lambda_g$ is a scalar controlling the volume of the ellipsoid, $\A_g$ is a diagonal matrix specifying the shape of the density contours, and $\D_g$ is an orthogonal matrix which determines the orientation of the corresponding ellipsoid \citep{Banfield:Raftery:1993, Celeux:Govaert:1995}. \citet[Table~1]{Fraley:Raftery:2006Mclust} report some parametrizations of within-group covariance matrices available in the \texttt{MCLUST} software, and the corresponding geometric characteristics.
Maximum likelihood estimates for this type of mixture models can be computed via the EM algorithm \citep{Dempster:Laird:Rubin:1977, Fraley:Raftery:2002}, while model selection could be based on the Bayesian information criterion (BIC) \citep{Fraley:Raftery:1998} or the integrated complete-data likelihood (ICL) criterion \citep{Biernacki:Celeux:Govaert:2000}.

In this paper we propose a dimension reduction method for model-based clustering, where we pursue the estimation of a reduced subspace which captures most of the clustering structure contained in the data. Following the work of \cite{Li:1991, Li:2000} on Sliced Inverse Regression (SIR), information on the dimension reduction subspace is obtained from the variation on group means and, depending on the estimated mixture model, on the variation on group covariances.
The estimated directions, ordered by importance as quantified by the associated eigenvalues, span a dimension reduction subspace. Observations may then be plotted on such a subspace, hence providing summary plots which visualize the clustering structure, and may help to understand the clustering and improve accuracy.

In Section~\ref{drmbc} we introduce the problem of clustering on a dimension reduced subspace, and we show that assignment of an observation to a given cluster is unchanged if performed on a suitable subspace.
Then we discuss estimation of the directions which span the reduced subspace, and we provide some of their properties. In Section~\ref{sel} we review and adapt a greedy search procedure for selecting a subset of the new constructed variables which retains most of the clustering information contained in the data. 
In Section~\ref{simul} we describe a data analysis example based on a synthetic data set; then we discuss results from simulation studies where we compare the proposed approach to model-based clustering performed on both the original variables and the leading principal components. 
Section~\ref{examples} presents applications of the proposed procedure on some real data sets. The final section contains some concluding remarks.

\section{Dimension reduction for model-based clustering}
\label{drmbc}

Classical procedures for dimensionality reduction are principal components analysis and factor analysis, both of which reduce dimensionality by forming linear combinations of the features. The first method seeks a lower-dimensional representation that account for the most variance of the features, while the second looks for the most correlation among the features. However, neither method correctly addresses the problem of visualizing any potential clustering structure.

\begin{figure}[htb]
\centering
\includegraphics[width=0.8\linewidth]{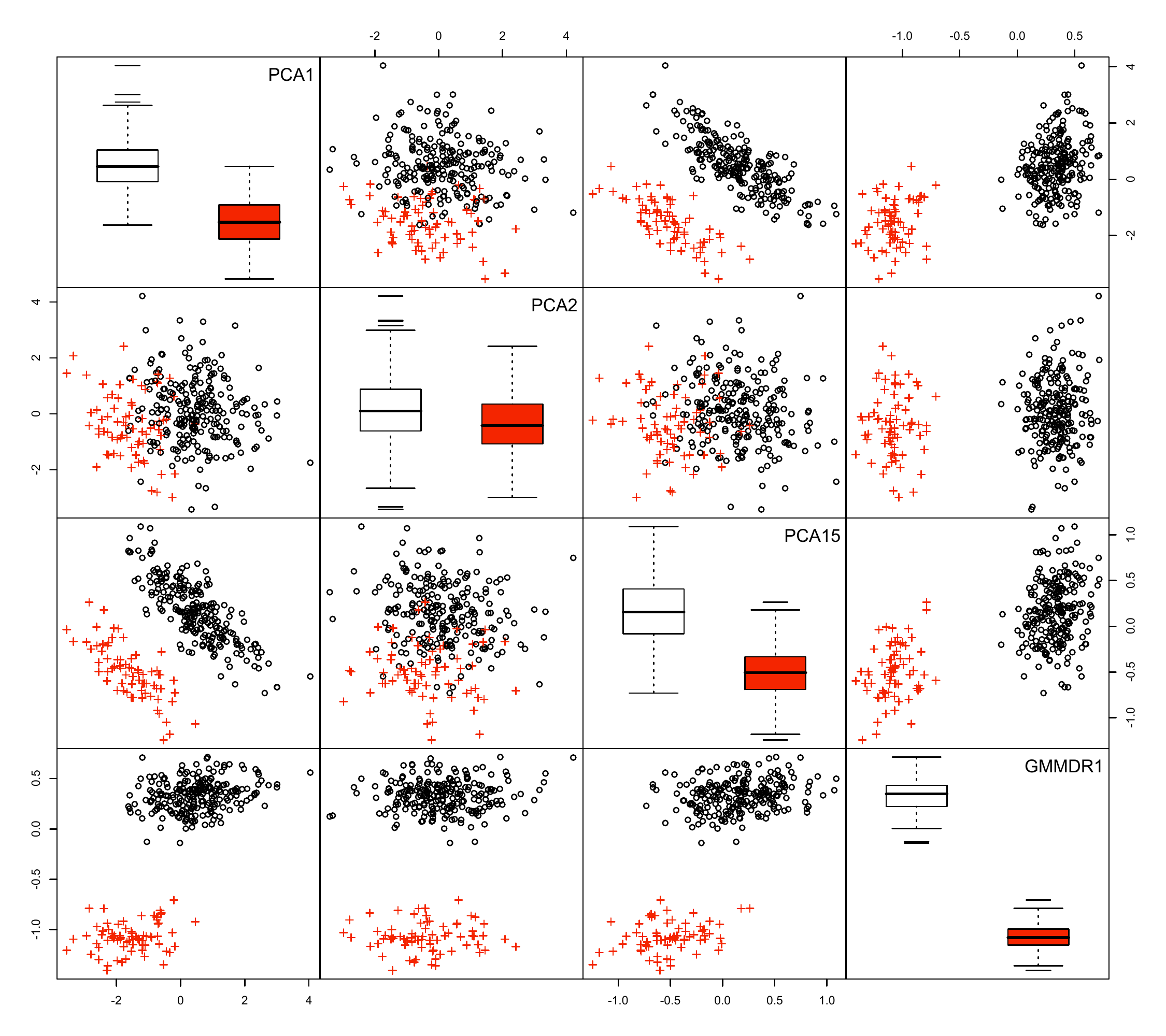}
\caption{Scatterplot matrix of 1st, 2nd and 15th PC, and 1st GMMDR variable for the Chang data with points marked according to cluster membership.}
\label{fig1:chang83}
\end{figure}

\citet{Chang:1983} discussed a simulated example from a 15-dimensional mixture model to show the failure of principal components as a method for reducing the dimension of the data before clustering. 
Let 
$\X = 0.5 \times d + d \times Y + \Z$, where
$d = 0.95 - 0.05 i$ $(i=1,\ldots,15)$,
$Y \sim \mathrm{Bernoulli}(0.2)$, 
$\Z \sim \mathrm{N}(\mub, \Sigmab)$ with mean
$\mub_{15\times 1} = [0,\ldots,0]\T$ and covariance matrix
$\Sigmab_{15\times 15} = [\sigma_{ij}]$, 
$\sigma_{ii}=1, \sigma_{ij}=-0.13 f_i f_j$, where the first 8 elements of $f$ are $-0.9$ and the last 7 are $0.5$. With this scheme the first 8 variables can be considered roughly as a block of variables with the same correlations, while the rest of the variables form another block. 
Using this scheme we simulated $n = 300$ data points.
Figure~\ref{fig1:chang83} shows the scatterplot matrix of the first, second and last principal components, and the first GMMDR variable (to be discussed on Section~\ref{estimation}) obtained from a two components EEE mixture model. As it can be seen, the first two PCs are not able to show any clustering information, but, as discussed by \citet{Chang:1983}, the first and the last PCs are needed. On the contrary, only one GMMDR variable is required to clearly separate the clusters. Furthermore, the coefficients (under unit norm constraint) of the linear combination defining such a direction reproduce the blocking structure used to simulate the variables, with the first 8 variables having coefficients approximately equal to $-0.32$ and the remaining 7 coefficients approximately equal to $0.16$.

\begin{figure}[htb]
\centering
\includegraphics[width=0.6\linewidth]{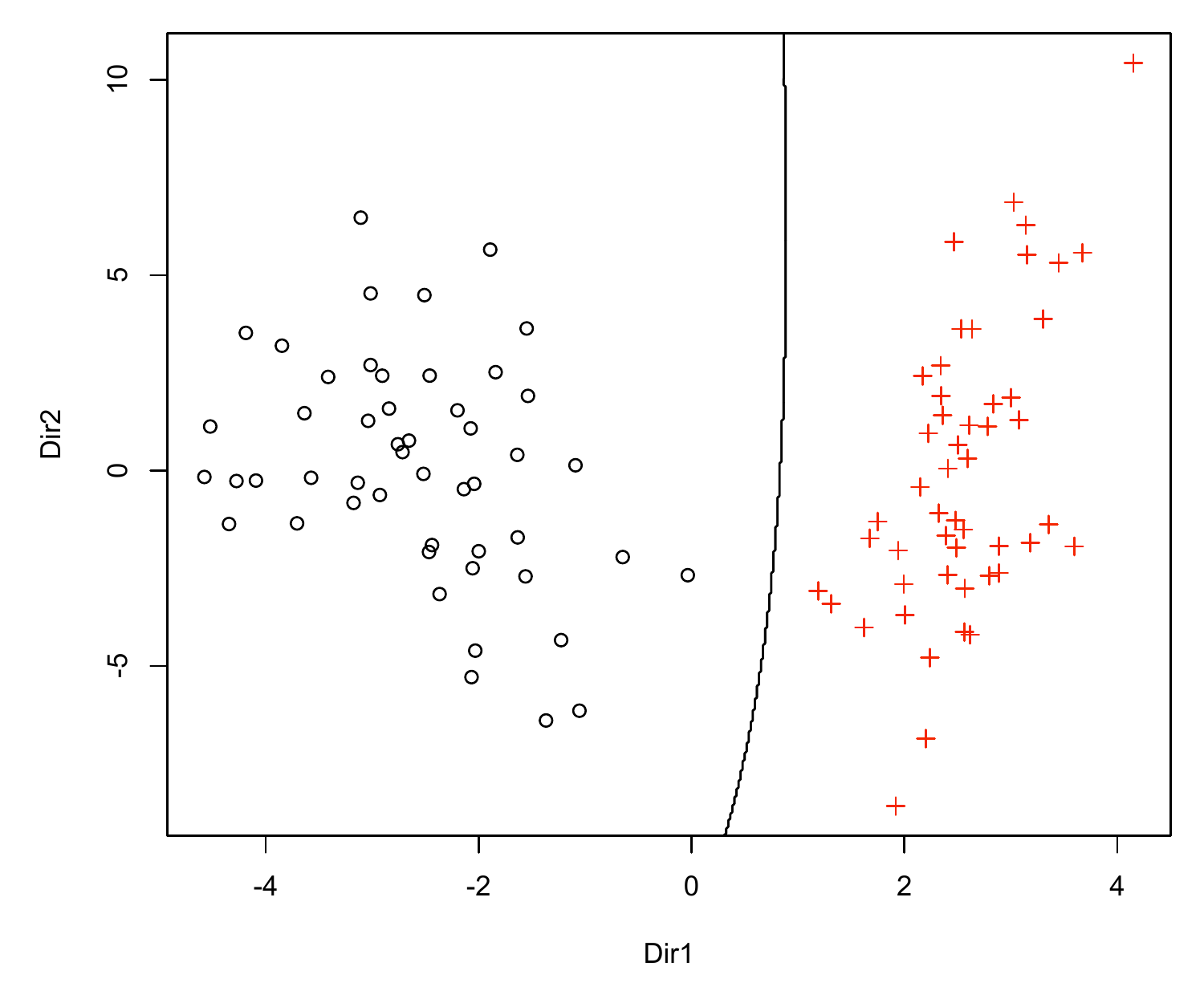}
\caption{Plot of \citet[p. 239--240]{McLachlan:Peel:2000} synthetic data  projected along the first two GMMDR directions, with points marked according to true group origin. The curve represents clustering boundary.}
\label{fig1:2cl5dim2groups}
\end{figure}

\cite{Tipping:Bishop:1999b} proposed a probabilistic principal component analysis based on a factor analysis model for the data assuming that the distribution of the errors are isotropic, i.e. the covariance matrix is proportional to the identity matrix. They also developed a mixture of probabilistic principal components analyzers model \citep{Tipping:Bishop:1999a}, and a hierarchical visualization algorithm for exploring clusters \citep{Bishop:Tipping:1998}.
\cite{McLachlan:Peel:2000} and \cite{McLachlan:Peel:Bean:2003} used mixtures of factor analyzers, by postulating that the distribution of the data within any group or latent class follows a factor analysis model. Parsimonious Gaussian mixture models based on mixtures of factor analyzers were discussed by \cite{McNicholas:Murphy:2008}. In order to deal with high dimensional data, \cite{Bouveyron:Girard:Schmid:2007} proposed a family of parsimonious GMMs on clustering subspaces.
All such models mainly address the problem of reducing the number of parameters to be estimated by a suitable reparametrization of the covariance matrix. Since they produce estimates of (latent) factor analyzers, they can also be used to reduce dimensionality. However, as noted by \citet[p. 248]{McLachlan:Peel:2000}, the resulting plots are not always useful in showing clustering structure. To support such claim, they presented an example using a simple two clusters synthetic data set where the differences between the group means are only in one variable, and with within-group variation in the other variables being relatively large. They showed that both principal components and factor analyzers were unable to show the underlying clustering structure
\citep[see][p. 240, 254--255]{McLachlan:Peel:2000}. We applied the GMMDR method proposed in this paper, and the data projected onto the subspace spanned by the first two directions are shown in Figure~\ref{fig1:2cl5dim2groups}: the two clusters appear well separated with one group showing less variation than the other.

\subsection{Clustering on a dimension reduced subspace}

Consider the $(p \times 1)$ vector $\X$ of random variables, and the discrete random variable $Y$ taking $G$ distinct values to indicate the $G$ clusters. Let $\betab$ denote a fixed $(p \times d)$ matrix, where $d \le p$, such that $Y \ind \X | \betab\T\X$.
This conditional independence statement says that the distribution of $Y|\X$ is the same as that of $Y|\betab\T\X$ for all values of $\X$ in its marginal sample space. As a consequence, the $(p \times 1)$ vector $\X$ can be replaced by the $(d \times 1)$ vector $\betab\T\X$ without loss of clustering information. If $d < p$, we have effectively reduced the dimension of the predictor vector. The matrix $\betab$ provides a basis for the subspace $\Space(\betab)$, which in a regression context is called a dimension-reduction subspace (DRS) for the regression of $Y$ on $\X$ \citep{Li:1991}. It can be shown that a minimum DRS may not be unique, but when several of such subspaces exist they all have the same dimension. Let $\Space_{Y|\X}$ denote the intersection of all DRSs. Under various reasonable conditions given in \citet[Chap.~6]{Cook:1998}, $\Space_{Y|\X}$ is a unique minimum DRS and it is called central DRS.
%\textit{**Rew.1 explicitly list the conditions in Cook (1998)**}

The assumption $Y \ind \X | \betab\T\X$ implies that $\Pr(Y=g|\X) = \Pr(Y=g|\betab\T\X)$, so we may write the density for the $g$-th component of the mixture model as
\begin{equation}
\begin{split}
f_g(\x) & = \frac{\Pr(Y=g|\x)f_{\X}(\x)}{\Pr(Y=g)} \\
        & = \frac{\Pr(Y=g|\betab\T\x)f_{\X}(\x)}{\Pr(Y=g)} \\
        & = f_g(\betab\T\x) \frac{f_{\X}(\x)}{f_{\betab\T\X}(\betab\T\x)},
\end{split}
\label{mixcomp}
\end{equation}
and for any two groups, $g$ and $k$, the ratio of the densities is
\begin{equation*}
\frac{f_g(\x)}{f_k(\x)} = \frac{f_g(\betab\T\x)}{f_k(\betab\T\x)}.
\end{equation*}
Thus, if $Y \ind \X | \betab\T\X$ the ratio of the conditional densities is the same whether it is computed on the original variables space or on the DRS, so the clustering information is completely contained in $\Space(\betab)$.
For instance, consider the simple case of two clusters with bivariate Gaussian distribution $\x|(Y=j) \sim N(\mub_j, \I_2)$ for $j=1,2$, where $\mub_1=[-1,0]\T$, $\mub_2=[1,0]\T$, and $\I_2$ is the $(2 \times 2)$ identity matrix. 
The ratio of the densities is thus given by
$$
\frac{f_1(\x)}{f_2(\x)} = \exp\{ -\frac{1}{2}(\x-\mub_1)\T(\x-\mub_1) + \frac{1}{2}(\x-\mub_2)\T(\x-\mub_2) \} = \exp\{-2x_1\}.
$$
It is straightforward to see that the direction containing all the clustering information is $\betab = [1,0]\T$, so the ratio of the densities projected along such direction is 
\begin{align*}
\frac{f_1(\betab\T\x)}{f_2(\betab\T\x)} & = \exp\{ -\frac{1}{2}(\betab\T\x-\betab\T\mub_1)^2 + \frac{1}{2}(\betab\T\x-\betab\T\mub_2)^2 \} \\
 & = \exp\{ -\frac{1}{2}(x_1+1)^2 + \frac{1}{2}(x_1-1)^2 \} = \exp\{-2x_1\}.
\end{align*}

In model-based clustering an observation is assigned to cluster $g$ by the maximum a posteriori principle (MAP), i.e. to the cluster for which the conditional probability given the data is a maximum: 
\begin{equation*}
\arg_g\max\Pr(Y=g|\x) = \frac{\pi_g f_g(\x)}{\sum_{j=1}^G \pi_j f_j(\x)}.
\end{equation*}
By equation \eqref{mixcomp}, the last expression is equivalent to
\begin{equation*}
\arg_g\max\Pr(Y=g|\betab\T\x) = \frac{\pi_g f_g(\betab\T\x)}{\sum_{j=1}^G \pi_j f_j(\betab\T\x)},
\end{equation*}
so the assignment of an observation to a cluster is unchanged if performed on the DRS. 
\citet[p. 156]{Cook:Yin:2001} considered $D(Y|\X)=\arg_g\max\Pr(Y=g|\x)$ and defined a discriminant subspace as a subspace $\Space$ such that $D(Y|\X) = D(Y|P_\Space\X)$ for all values of $\X$, where $P_\Space$ denotes the projection operator onto $\Space$ with respect to the usual inner product.
The intersection of all discriminant subspaces, if it exists, is itself a discriminant subspace, and is called the central discriminant subspace (CDS) and denoted by $\Space_{D(Y|\X)}$. Finally, the CDS is a subset of the central DRS, $\Space_{D(Y|\X)} \subseteq \Space_{Y|\X}$, and in some cases may be a proper subset.

In the case of known group memberships, a variety of methods for dimensionality reduction have been proposed \citep{Cook:Yin:2001}. These methods, such as Sliced Inverse Regression (SIR) and Sliced Average Variance Estimation (SAVE), are not restricted to any particular classification method.
%Here we address a different problem: provide a dimension reduction and visualization method for a model-based clustering procedure (where group memberships are unknown).
In the following subsection we discuss estimation of a dimension reduction subspace for a model-based clustering procedure.

\subsection{Estimation of GMMDR directions}
\label{estimation}

Suppose we describe a set of $n$ observations on $p$ variables through a $G$ components Gaussian mixture model (GMM) of the form
$f(\x) = \sum_{g=1}^G \pi_g \phi(\x|\mub_g, \Sigmab_g)$.

From a graphical point of view, we pursue the smallest subspace that can capture the clustering information contained in the data. A natural starting point would require to look for those directions where the cluster means $\mub_g$ vary as much as possible, provided each direction is orthogonal to the others.
This amounts to solving the following optimization problem:
$$
\arg_{\betab}\max \betab\T \Sigmab_{B} \betab,
\quad\text{subject to}\quad
\betab\T \Sigmab \betab = \I_d,
$$
where $\Sigmab_B = \sum_{g=1}^G \pi_g (\mub_g - \mub)(\mub_g - \mub)\T$ is the between-cluster covariance matrix,
$\Sigmab = n\inv \sum_{i=1}^n \\ (\x_i - \mub)(\x_i - \mub)\T$ is the covariance matrix with $\mub = \sum_{g=1}^G \pi_g \mub_g$, 
$\betab \in \Real^{p \times d}$ is the spanning matrix, and
$\I_d$ is the $(d \times d)$ identity matrix. The solution to this constrained optmization is given by the eigendecomposition of the kernel matrix $\M_I \equiv \Sigmab_B$ with respect to $\Sigmab$. The eigenvectors, corresponding to the first $d$ largest eigenvalues, $[\v_1,\ldots,\v_d] \equiv \betab$, provide a basis for the subspace $\Space(\betab)$ which shows the maximal variation among cluster means. There are at most $d=\min(p,G-1)$ directions which span this subspace.

This procedure is similar to the sliced inverse regression (SIR) algorithm \citep{Li:1991}. SIR is a dimension reduction method which exploits the information from the inverse mean function \cite[see][Chap.~2]{Li:2000}. Instead of conditioning on the response variable (or a sliced version of it if continuous), 
here we condition on the cluster memberships. SIR directions span at least part of the dimension reduction subspace \citep[Chap.~6]{Cook:1998} and they provide an intuitive and useful basis for constructing summary plots. However, they may miss relevant structures in the data when within-cluster covariances are different.

The SIR$_{II}$ method \citep[Chap.~5]{Li:2000} exploits the information coming from the differences in the class covariance matrices. The kernel matrix is now defined as:
\begin{equation*}
\M_{II} = \sum_{g=1}^G \pi_g (\Sigmab_g - \Sigmabar) \Sigmab\inv (\Sigmab_g - \Sigmabar)\T,
\end{equation*}
where $\Sigmabar = \sum_{g=1}^G \pi_g \Sigmab_g$ is the pooled within-cluster covariance matrix, and directions are found through the eigendecomposition of $\M_{II}$ with respect to $\Sigmab$.
Albeit the corresponding directions show differences in group covariances, they are usually not able to also show location differences.

The proposed approach aims at finding those directions which, depending on the selected Gaussian mixture model, are able to display both variation in cluster means and variation in cluster covariances.

\begin{definition}
Consider the following kernel matrix
\begin{equation}
\M = \M_I \Sigmab\inv \M_I + \M_{II}.
\label{kernel}
\end{equation}
The basis of the dimension reduction subspace $\Space(\betab)$ is the solution of the following constrained optimization:
\begin{equation*}
\arg_{\betab}\max \betab\T \M \betab, 
\quad\text{subject to}\quad
\betab\T \Sigmab \betab = \I_d.
\end{equation*}
This is solved through the generalized eigendecomposition
\begin{equation}
\begin{split}
\M\v_i = l_i\Sigmab\v_i, & \quad \v_i\T \Sigmab \v_j = 1 \;\text{ if } i = j\text{, and } 0 \text{ otherwise}, \\
                        & \quad l_1 \ge l_2 \ge \ldots \ge l_d > 0.
\end{split}
\label{gmmdr:decomp}
\end{equation}
Thus, the basis of the required subspace is given by the GMMDR directions $\betab \equiv [\v_1,\ldots,\v_d]$.
\end{definition}

The kernel matrix \eqref{kernel} contains information from variation on both cluster means and cluster covariances. For mixture models which assume constant within-cluster covariance matrices, such as models E, EII, EEI, and EEE \citep[see][Table~1]{Fraley:Raftery:2006Mclust}, the subspace spanned by $\M$ is equivalent to that spanned by $\M_I$.

\begin{remark}
Let $\Space(\betab)$ be the subspace spanned by the $(p \times d)$ matrix $\betab$ of eigenvectors from \eqref{gmmdr:decomp}, and $\mub_g$ and $\Sigmab_g$ be the means and covariance, respectively, for the $g$-th cluster. It can be easily seen that the projection of the parameters onto the subspace $\Space(\betab)$ are given by $\betab\T\mub_g$ and $\betab\T\Sigmab_g\betab$. Furthermore, the projection of the $(n \times p)$ data matrix $\X$ onto $\Space(\betab)$ can be computed as $\Z=\X\betab$; we call these GMMDR variables.
\end{remark}

\begin{remark}
The raw coefficients of the estimated directions are uniquely determined up to multiplication by a scalar, whereas the associated directions from the origin are unique. Hence, we can adjust their length such that they have unit length, i.e. each direction is normalized as $\betab_j \equiv \v_j/||\v_j||$ for $j=1,\ldots,d$. In matrix form, let $\D = \diag(\V\T\V)$, where $\V$ is the matrix of eigenvectors from \eqref{gmmdr:decomp}, then $\betab \equiv \V \D^{-1/2}$. 
For the GMMDR variables
$$
\Cov(\Z) = \betab\T \Sigmab \betab = \D^{-1/2} \V\T \Sigmab \V \D^{-1/2} = \D\inv = \diag(1/||\v_j||^2).
$$
Thus, they are uncorrelated, and with variances inversely related to the squared length of the eigenvectors of the kernel matrix, while GMMDR directions are orthogonal with respect to $\Sigmab$-inner product.
\end{remark}

We now provide some properties as propositions whose proofs are reported in the Appendix \ref{appendix:proofs}.

% Affine transformation Invariance
\begin{proposition}
\label{prop:invariance}
GMMDR directions are invariant under affine transformation $\x  \mapsto \C \x + \a$, for $\C$ a nonsingular matrix and $\a$ a vector of real values. Thus, GMMDR directions in the transformed scale are given by $\C\inv\betab$.
\end{proposition}

% Interpretation of eigenvalues
\begin{proposition}
\label{prop:eigenvalues}
Each eigenvalue of the eigendecomposition in \eqref{gmmdr:decomp} can be decomposed in the sum of the contributions given by the squared variance of the between group means and the average of the squared within group variances along the corresponding direction of the projection subspace, i.e.
\begin{equation*}
l_i = \Var(\Exp(Z_i|Y))^2 + \Exp(\Var(Z_i|Y)^2) 
\qquad \mathrm{for}\; i=1,\ldots,d.
\end{equation*}
\end{proposition}
Based on this result we may provide an interpretation to the contribution of each direction to the visualization of the clustering structure. Furthermore, directions corresponding to small eigenvalues provide little or no information about differences in cluster means or covariances.

\begin{remark}
Let $\X$ be the $(n \times p)$ sample data matrix which we assume, with no loss of generality, to have zero mean column-vectors. 
The sample version $\hat{\M}$ of \eqref{kernel} is obtained using the corresponding estimates from the fit of a Gaussian mixture model via the EM algorithm. Then, GMMDR directions are calculated from the generalized eigen-decomposition of $\hat{\M}$ with respect to $\hat{\Sigmab}$.
\end{remark}

\begin{remark}
The dimension of the estimated subspace $\Space({\betab})$ is $d = \min\{p, G-1\}$ for models which assume equal within-cluster covariance matrices (i.e. models E, EII, EEI, and EEE), while $d \le p$ otherwise. The GMMDR directions are ordered based on the associated eigenvalues, so those directions corresponding to approximately zero eigenvalues can be discarded in practice since clusters largely overlap along such directions. A more formal approach to the selection of relevant directions is discussed in Section~\ref{sel}.

\end{remark}

\section{Subset selection of GMMDR variables}
\label{sel}

In the previous section we have discussed the estimation of GMMDR variables, which is basically a linear mapping method onto a suitable subspace. This can be viewed as a form of (soft) feature extraction, where the components are reduced through a set of linear combinations of the original variables. However, it may be possible that a subset of the estimated GMMDR variables provide either no clustering information or redundant clustering information already provided by other variables.

Consider the partitioned $(p \times p)$ matrix $(\betab,\betab_0)$, which forms a basis for $\Real^p$. Based on proposition 3 of Cook and Yin (2000, p.157), if $\betab_0\T\X \ind \betab\T\X | Y$ and $\betab_0\T\X \ind Y$ then $\Space(\betab)$ is a DRS. This is useful since, albeit it does not indicate how to construct $\betab$, it indicates that unnecessary variables $\betab_0\T\X$ are independent from relevant variables $\betab\T\X$ within groups, and that they are also marginally independent from $Y$. 
In our case, given that unnecessary GMMDR variables provide no clustering information but require parameters estimation, we would like to detect and discard them.

In the following we consider a subset selection method for GMMDR variables following the approach of \citet{Raftery:Dean:2006}, who proposed the use of BIC to approximate Bayes factors for comparing mixture models fitted on nested subsets of variables. A generalization of this approach has been recently discussed by \citet{Maugis:etal:2009}.

\subsection{A criterion for feature selection}
\label{criterion}

Let $\Set$ be the set of indices with $\dim(\Set)=q$ indexing a subset of $q$ features from the original $d$ GMMDR variables $Z$ ($q \le d$), and $\Set' = \{\Set\setminus i\} \subset \Set$ be the set of $\dim(\Set')=q-1$ obtained excluding the $i$-th feature index from $\Set$.
The comparison of the two nested subsets can be recast as a model comparison problem and addressed using BIC difference as an approximation to Bayes factor. Following \citet{Raftery:Dean:2006}, the BIC difference for the inclusion of the $i$-th feature is given by
\begin{equation}
\BICdiff(Z_{i \in \Set}) = \BICclust(Z_\Set) - \BICnotclust(Z_{\Set}),
\label{BICdif}
\end{equation}
where $\BICclust(Z_\Set)$ is the BIC value for the ``best'' model fitted using features in $\Set$, and $\BICnotclust(Z_{\Set})$ is the BIC value for no clustering. The latter can be written as 
\begin{equation*}
\BICnotclust(Z_{\Set}) = \BICclust(Z_{\Set'}) + \BICreg(Z_i|Z_{\Set'}),
\end{equation*}
i.e. the BIC value for the ``best'' model fitted using features in $\Set'$ plus the BIC value for the regression of the $i$-th feature on the remaining $(q-1)$ features in $\Set'$. Noting that GMMDR variables are orthogonal, the general formula for $\BICreg$ \citep[see][eq. (7)]{Raftery:Dean:2006} reduces to
$\BICreg(Z_i|Z_{\Set'}) = -n\log(2\pi) - n\log(\hat{\sigma}_i^2) - n - (q+1)\log(n)$,
where $\hat{\sigma}_i^2$ is the maximum likelihood estimate of the $i$-th feature variance. 
In all cases the ``best'' clustering model is identified with respect to both the number of mixture components and model parametrization.

\subsection{A greedy search algorithm for selecting the ``best'' subset}
\label{greedy}
GMMDR variables are defined as orthogonal linear combinations of the original variables, and it is likely that only a subset of them is needed for clustering.
However, the space of all possible subsets of size $q$, with $q$ ranging from 1 to $d$, has number of elements equal to $2^d-1$. An exhaustive search soon becomes unfeasible, even for moderate values of $d$. To alleviate this problem, \citet{Raftery:Dean:2006} proposed a greedy search algorithm for finding a local optimum in the model space. The search has both a forward step which evaluates inclusion of a proposed variable, and a backward step which evaluates exclusion of one of the currently included variables. The algorithm stops when consecutive inclusion and exclusion steps are rejected. Since GMMDR variables are orthogonal the search can be simplified by skipping the backward step. This allows us to considerably speed up the computing time.

The search starts by selecting the single GMMDR variable which maximizes the BIC criterion. This is equivalent to using the statistic in equation \eqref{BICdif}, with $\Set'=\emptyset$ and $\Set=\{i\}$ for any $i=1,\ldots,d$. 
Thus, at the first step equation \eqref{BICdif} measures the difference between the best clustering model and the single cluster model for each variable.
In the following steps, we select the GMMDR variate which maximizes the BIC difference \eqref{BICdif}, so taking into account the features already included. We keep adding GMMDR variates until the BIC difference becomes negative. 
A detailed description of the greedy search algorithm is reported in Appendix \ref{appendix:greedysearch}.

At each step, the search over the model space is usually performed with respect to both the model parametrization and the number of clusters. However, we may want to fix the model parameters at the estimated values obtained using the full set of features; in such a case, the order in which the GMMDR directions are selected follows the ordering of the associated eigenvalues.

\subsection{An iterative procedure for feature selection}

The greedy search discussed in the previous section is likely to discard some of the GMMDR variates, in particular those associated with small eigenvalues which do not carry any clustering information once other features have been included. A GMM may then be fitted on such constructed variables and the corresponding GMMDR directions estimated. 
The feature selection step can then be repeated, and the whole process iterated until no directions could be dropped. Depending on the number of initial features, the number of iterations required is usually small. The whole process is described in the following algorithm:
\begin{center}
\begin{minipage}{0.9\linewidth}
\hrule\vspace*{1mm}
\textbf{Algorithm}: GMMDR estimation and feature selection process
\vspace*{1mm}\hrule
\begin{enumerate}
\item Fit a GMM on the original data.
\item Estimate GMMDR directions using the method in Section~\ref{estimation} and project the data onto the estimated subspace.
\item Perform feature selection using the greedy search discussed in Section~\ref{greedy}.
\item Fit a GMM on the selected GMMDR variables and go to step 2.
\item Repeat steps 2--4 until none of the features could be dropped.
\end{enumerate}
\hrule
\end{minipage}
\end{center}

\section{Simulations}
\label{simul} 

In this section we present a data analysis example based on a synthetic data set to describe the proposed approach, followed by some simulation studies. 
The adjusted Rand index (ARI), proposed by \citet{Hubert:Arabie:1985}, is adopted for evaluating the clustering obtained and comparing the results arising from competing mixture models. The ARI gives a measure of agreement between two partitions, one estimated by a statistical procedure independent of the labelling of the groups, and one being the true classification. This index has zero expected value in the case of random partition, and it is bounded above by 1. Thus, higher values represent a better performance.

\subsection{Synthetic data on three overlapping clusters with different shapes}
\label{simul1} 

Consider a simulated data set with three overlapping clusters on ten variables, with each cluster sample size equal to $n_g=50$ ($g=1,2,3$). The first three variables have been generated from a multivariate normal distribution with means 
$\mub_1 = [0,0,0]\T$,
$\mub_2 = [4,-2,6]\T$,
$\mub_3 = [-2,-4,2]\T$,
and within-groups covariances 
\begin{displaymath}
\Sigmab_1 = \begin{bmatrix}
                   1 & 0.9 & 0.9 \\
                   0.9 & 1 & 0.9 \\
                   0.9 & 0.9 & 1 \\
\end{bmatrix}
,\;
\Sigmab_2 = \begin{bmatrix}
                   2 & -1.8 & -1.8 \\
                   -1.8 & 2 & -1.8 \\
                   -1.8 & -1.8 & 2 \\
\end{bmatrix}
,\;\text{and }
%\end{displaymath}
%\begin{displaymath}
\Sigmab_3 = \begin{bmatrix}
                   0.5 & 0 & 0 \\
                   0 & 0.5 & 0 \\
                   0 & 0 & 0.5 \\
\end{bmatrix}.
\end{displaymath}
Seven noise variables, generated independently from a standard normal distribution, were also added. Thus, clustering information is only contained in the first three variables, with the remaining ones been simply noise.
The correct model we hope to find is the VVV model with $G=3$.

The GMM with the largest BIC is the seven clusters EEI model, closely followed by the model with six clusters. Both models have the wrong number of components, thus yielding a poor classification accuracy. Constraining $G=3$ the selected model improves the ARI, albeit there are still some misclassified data points (see Table \ref{tab1:3cl10dim}). 
Using the ``best'' unconstrained GMM we obtained the GMMDR directions; there are $d=6$ of such directions, but only the first three are retained by the feature selection step. 
As it can be seen from Table~\ref{tab1:3cl10dim}, the final VVV mixture model with 3 clusters fitted on the three selected GMMDR variables is able to recover the original clustering structure. 

\begin{table}[htb]
\centering
\caption{Clustering results for the three clusters simulated data.}
\label{tab1:3cl10dim}
\begin{tabular}{llrrr}
\hline\noalign{\smallskip}
Method & Model & $G$ & BIC & ARI \\
\noalign{\smallskip}\hline\noalign{\smallskip}
GMM (unconstrained) & EEI & 7 & -4731.56 & 0.6674 \\
GMM (constrained)   & VII & 3 & -4820.57 & 0.8671 \\
GMMDR (3 directions) & VVV & 3 & -1339.40 & 1.0000 \\
\noalign{\smallskip}\hline
\end{tabular}
\end{table}

\begin{figure}[htb]
\centering
\includegraphics[width=0.43\linewidth]{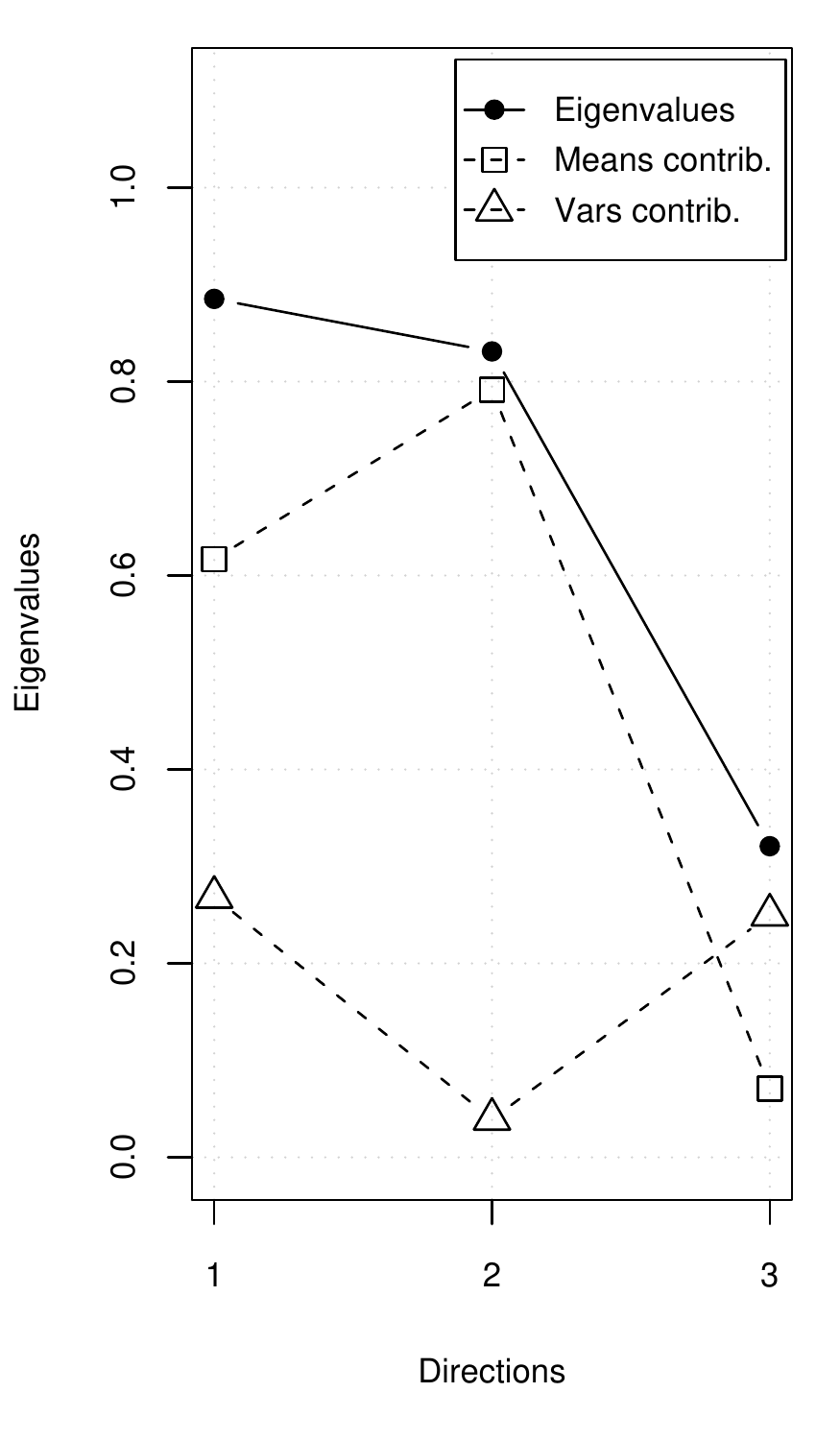}\;
\includegraphics[width=0.49\linewidth]{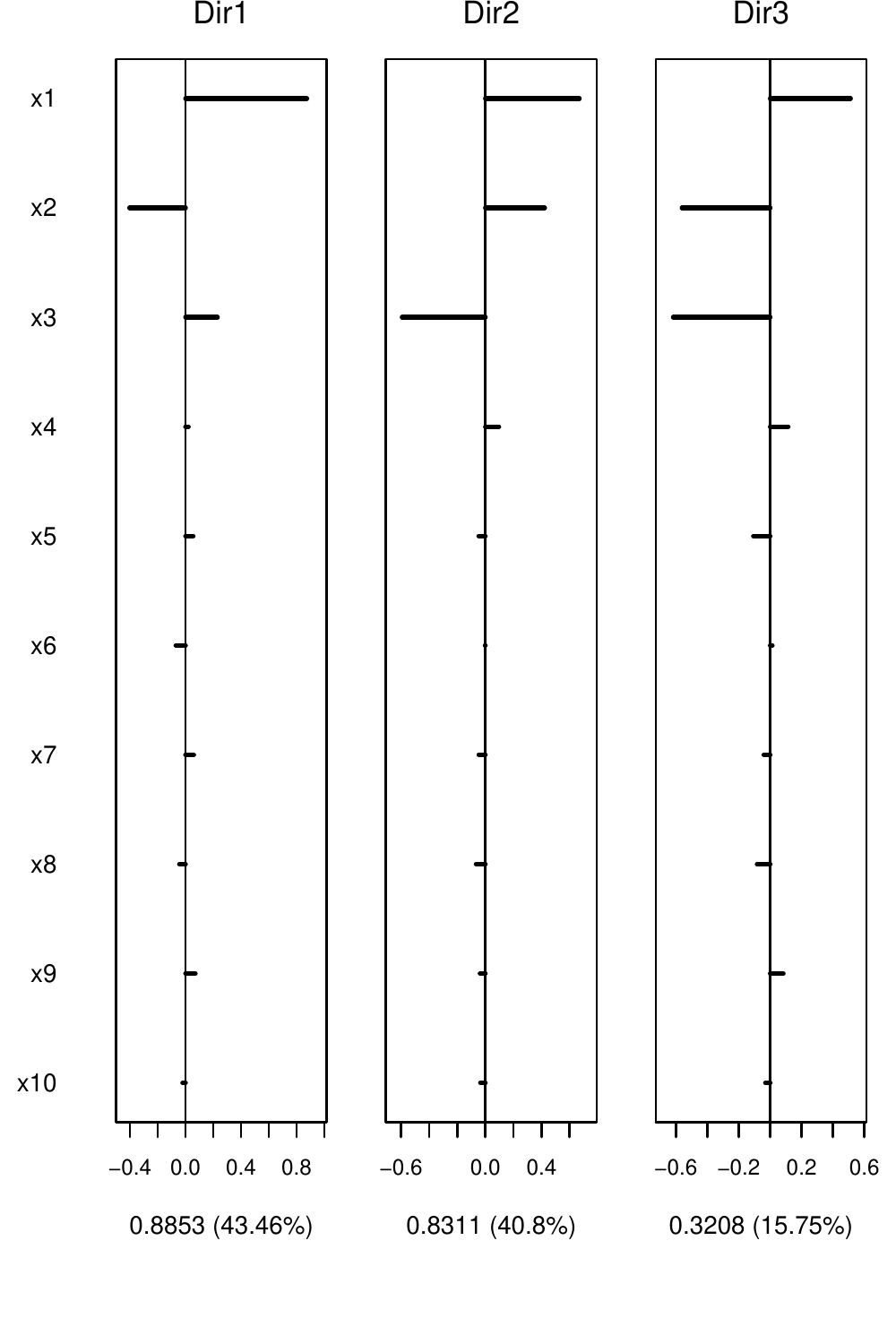}
\caption{Plot of eigenvalues for each estimated direction with contributions from means and variances differences among clusters (left panel) and coefficients defining the GMMDR directions (right panel).}
\label{fig1:3cl10dim_evalues_coefs}
\end{figure}

The left panel of Figure~\ref{fig1:3cl10dim_evalues_coefs} shows the eigenvalues associated with the GMMDR directions. From such a plot we can see that the first two directions are the most important, with the first showing both difference in means and variances among clusters, while the second almost only differences in means. The last direction adds a small amount of clustering information, mostly from differences in variances.
The coefficients for the selected GMMDR directions (see the right panel of Figure~\ref{fig1:3cl10dim_evalues_coefs}) indicate that only the first three variables seem to be required, since the coefficients for the remaining ones are almost zero. 

\begin{figure}[htb]
\centering
\includegraphics[width=0.9\linewidth]{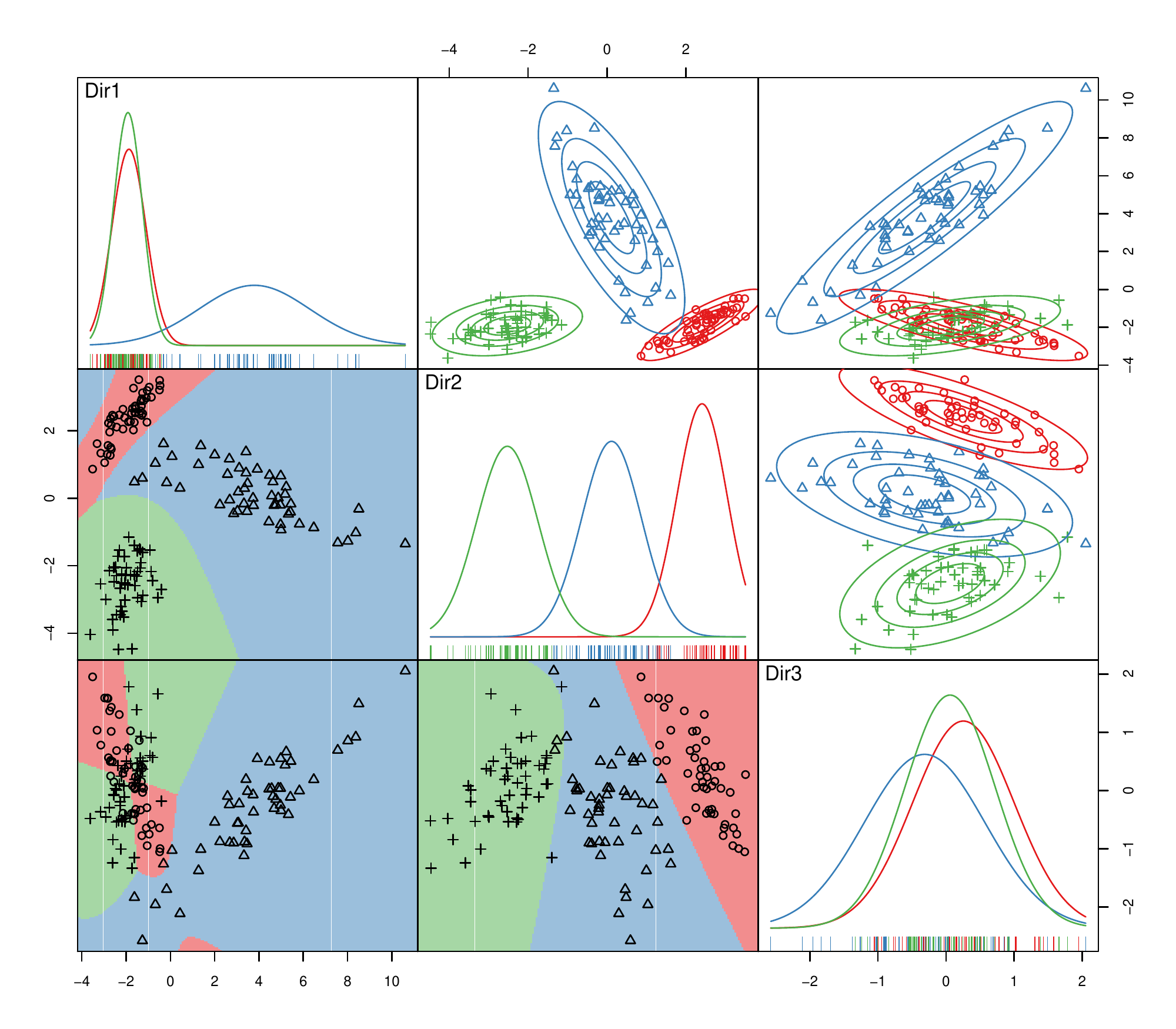}
\caption{Scatter plots of GMMDR variables for the three clusters simulated data: bivariate component density contours (above diagonal), marginal univariate component densities (diagonal), and MAP regions (below diagonal).}
\label{fig2:3cl10dim_pairs}
\end{figure}

Graphs contained in Figure~\ref{fig2:3cl10dim_pairs} use GMMDR directions to convey clustering information about the fitted GMM.
Plots along the diagonal show univariate component densities, while plots above the diagonal report contours of the estimated densities for each component.
Such graphs clearly reflect the characteristics of the mixture model fitted; in particular, clusters have different volume, shape and orientation. Furthermore, it is quite clear that clusters on the projection subspace appear to be well separated along the first two directions, as it can be seen from the maximum a posteriori (MAP) classification areas below the diagonal of Figure~\ref{fig2:3cl10dim_pairs}.
 
\subsection{Simulation studies}
\label{simul2} 

Here we discuss the results from some simulation studies we conducted to compare the proposed GMMDR approach to model-based clustering performed on both the original variables and the leading principal components. 
The data sets used for such comparison are generated from some overlapping Gaussian mixtures with different covariance matrix structures.
In addition, we investigated the effect of the inclusion of a certain number of redundant and noise variables. The former are variables correlated with those bringing clustering information, so they only provide redundant information. The latter are variables whose distribution do not depend on the cluster label, thus they only introduce some noise which tends to obscure the underlying clustering structure to be recovered. 

The simulation schemes used for generating the data are described below.

\paragraph{Model 1: three overlapping clusters with common covariances.}
% sim.data2

In the first simulation scheme each data set was generated from three overlapping clusters with equal mixing probabilities on three variables. These were generated from a multivariate normal distribution with means
$\mub_1 = [0,0,0]\T$,
$\mub_2 = [0,2,2]\T$,
$\mub_3 = [2,-2,-2]\T$,
and common covariance matrix
\begin{equation*}
\Sigmab = \left[\begin{matrix}
2.0 & 0.7 & 0.8 \\
0.7 & 0.5 & 0.3 \\
0.8 & 0.3 & 1.0 \\
\end{matrix}\right].
\end{equation*}
To this basic setup, which correponds to a EEE mixture model \citep[see][Table~1]{Fraley:Raftery:2006Mclust}, we added two further scenarios. In the first scenario (\textit{noise variables}), seven noise variables were generated from independent standard normal variables. In the second scenario (\textit{redundant and noise variables}), we generated three variables correlated with each clustering variable (with correlation coefficients equal to 0.9, 0.7, 0.5, respectively) and four independent standard normal variables. In both cases, there was a total of ten variables.

\paragraph{Model 2: three overlapping clusters with constant shapes.}
% sim.data12

In this simulation scheme we generated observations with equal prior from three classes using the mixture model VEV. The means for each class were 
$\mub_1 = [0,0,0]\T$, $\mub_2 = [4,-2,6]\T$, and $\mub_3 = [-2,-4,2]\T$, 
while for the covariance matrices $\Sigmab_g = \lambda_g\D_g\A\D_g\T$ ($g=1,2,3$) the scale, shape and orientation parameters were, respectively, $\lambda = [0.2,0.5,0.8]\T$, $\A = \diag(1,2,3)$, and 
\begin{displaymath}
\D_1 = \begin{bmatrix}
1 & 0.6 & 0.6 \\
0.6 & 1 & 0.6 \\
0.6 & 0.6 & 1 \\
\end{bmatrix}
,\;
\D_2 = \begin{bmatrix}
2 & -1.2 & 1.2 \\
-1.2 & 2 & -1.2 \\
1.2 & -1.2 & 2
\end{bmatrix}
,\;
\D_3 = \begin{bmatrix}
0.5 & 0 & 0 \\
0 & 0.5 & 0 \\
0 & 0 & 0.5 
\end{bmatrix}.
\end{displaymath}
As for the previous model setup, we also considered two further scenarios with the addition of noise variables in the first case, and noise and redundant variables in the second case.

\paragraph{Model 3: three overlapping clusters with unconstrained covariances.}
% sim.data5

For this last scheme we simulated each data set from three overlapping clusters with equal mixing probabilities on three variables. 
Clustering variables were generated from a multivariate normal distribution with parameters set as in the synthetic example of Section~\ref{simul1}, which correspond to a VVV parametrization with $G=3$.
Again, we also considered the effect of adding only noise variables, and noise and redundant variables.

\medskip

For each data set simulated as described above we fitted a GMM with number of clusters in the range $[1,15]$ and different model pa\-ra\-metrizations, then we selected the model with the highest BIC. A GMM was also fitted to the leading principal components computed from the correlation matrix, or equivalently using standardized variables; the number of components was chosen using the \textit{Kaiser's rule}, i.e. selecting those with eigenvalues larger than one \citep[p. 114--115]{Jolliffe:2002}. Finally, we applied the GMMDR approach discussed in Section~\ref{estimation}, followed by the subset selection procedure of Section~\ref{sel}. For each estimated mo\-del we computed the adjusted Rand index \citep{Hubert:Arabie:1985} as a criterion to evaluate the clustering partitions obtained: higher values of ARI correspond to better performance. 

The results of the simulation procedure based on 1000 repetitions for increasing sample sizes are shown on Table~\ref{tab1:sim}, where we reported the average adjusted Rand index for the three fitting procedures to be compared and the corresponding standard errors. 

When no noise variables are included, the GMMDR procedure yields equivalent clustering accuracy to GMM on the original variables. However, when irrelevant features (either noise or redundant ones) are present the GMMDR procedure significantly improves the accuracy as measured by the adjusted Rand index. 
This is particular evident for smaller sample sizes ($n=100, 300$) when noise variables are present, and when the underlying clusters have different within cluster covariance matrices.
The PCA+GMM procedure leads to uniformly small values of ARI, confirming that reducing the dimensionality of the problem through principal components does not help to discover clustering structures in the data.
As expected, as the sample size increases all the three methods improve in accuracy, with the GMM method on the original variables and the GMMDR approach leading essentially to the same accuracy for $n=1000$.

A further aspect was also checked, the number of clusters selected by each procedure (results not shown here). Recalling that for all the simulation schemes the number of clusters was three, GMMDR almost always selects the true value, whereas GMM tends to select a larger number of components when noise variables are included and the clusters have different covariance structures. PCA+GMM performs worst than the other two methods, often under-estimating the true number of clusters in the presence of noise variables and for smaller sample sizes ($n=100, 300$).  

\begin{table}[htb]
\centering
\caption{Average adjusted Rand index (with standard errors within parenthesis) based on 1000 simulations.}
\label{tab1:sim}
\smallskip\small
\setlength{\tabcolsep}{3pt}
%\renewcommand\arraystretch{0.8}
%\begin{tabular}{lp{0.2\linewidth}p{0.2\linewidth}p{0.35\linewidth}}
\begin{tabular}{l@{\hspace{10pt}}ccc@{\hspace{10pt}}ccc@{\hspace{10pt}}ccc}
\hline\noalign{\smallskip}
 & \multicolumn{3}{c}{\textit{No noise}} & \multicolumn{3}{c}{\textit{Noise variables}} & \multicolumn{3}{c}{\textit{Noise and redundant variables}} \\
\hline\noalign{\smallskip}
\textit{Model 1 (EEE)} & $n = 100$ & $n = 300$ & $n = 1000$ & $n = 100$ & $n = 300$ & $n = 1000$ & $n = 100$ & $n = 300$ & $n = 1000$ \\
\hline\noalign{\smallskip}
GMM     & 0.9684 & 0.9744  & 0.9754  & 0.6869 & 0.8918 & 0.9742 & 0.8518 & 0.9695  & 0.9743 \\ 
     & (0.0420) & (0.0166) & (0.0083) & (0.1363) & (0.1217) & (0.0090) & (0.1889) & (0.0241) & (0.0087) \\  
PCA+GMM & 0.8591 & 0.9105 & 0.9146 & 0.4989 & 0.8352 & 0.9091 & 0.5361 & 0.7518 & 0.7942 \\
& (0.1626) & (0.0446) & (0.0221) & (0.1700) & (0.1256) & (0.0298) & (0.1342) & (0.0909) & (0.0239) \\ 
GMMDR  & 0.9716 & 0.9742 & 0.9753 & 0.8612 & 0.9674 & 0.9742 & 0.8832 & 0.9699 & 0.9742 \\
 & (0.0347) & (0.0172) & (0.0088) & (0.1426) & (0.0234) & (0.0090) & (0.1613) & (0.0197) & (0.0088) \\ 
\hline\noalign{\smallskip}
\textit{Model 2 (VEV)} & $n = 100$ & $n = 300$ & $n = 1000$ & $n = 100$ & $n = 300$ & $n = 1000$ & $n = 100$ & $n = 300$ & $n = 1000$ \\
\hline\noalign{\smallskip}
GMM     & 0.9738 & 0.9802 & 0.9819 & 0.7316 & 0.6328 & 0.9808 & 0.6648 & 0.6811 & 0.9807 \\
           & (0.0302) & (0.0141) & (0.0073) & (0.0825) & (0.0793) & (0.0081) & (0.0913) & (0.1572) & (0.0078) \\
PCA+GMM & 0.6562 & 0.6789 & 0.7130 & 0.1907 & 0.4307 & 0.6458 & 0.4579 & 0.6465 & 0.8018 \\
		   & (0.2745) & (0.2138) & (0.0824) & (0.1386) & (0.1700) & (0.1417) & (0.1780) & (0.1483) & (0.0309) \\
GMMDR  & 0.9709 & 0.9802 & 0.9819 & 0.9201 & 0.9747 & 0.9806 & 0.9231 & 0.9727 & 0.9806 \\
           & (0.0351) & (0.0141) & (0.0073) & (0.0799) & (0.0165) & (0.0082) & (0.0811) & (0.0169) & (0.0077) \\
\hline\noalign{\smallskip}
\textit{Model 3 (VVV)} & $n = 100$ & $n = 300$ & $n = 1000$ & $n = 100$ & $n = 300$ & $n = 1000$ & $n = 100$ & $n = 300$ & $n = 1000$ \\
\hline\noalign{\smallskip}
GMM     & 0.9960 & 0.9981 & 0.9983 & 0.6937 & 0.8877 & 0.9751 & 0.8434 & 0.9708 & 0.9742 \\
           & (0.0115) & (0.0043) & (0.0024) & (0.1260) & (0.1189) & (0.0081) & (0.1932) & (0.0194) & (0.0092) \\
PCA+GMM & 0.9805 & 0.9896 & 0.9899 & 0.4917 & 0.8357 & 0.9074 & 0.5396 & 0.7493 & 0.7945 \\
           & (0.0534) & (0.0105) & (0.0053) & (0.1663) & (0.1277) & (0.0316) & (0.1353) & (0.0984) & (0.0236) \\
GMMDR  & 0.9952 & 0.9981 & 0.9983 & 0.8643 & 0.9674 & 0.9751 & 0.8799 & 0.9712 & 0.9740  \\
           & (0.0146) & (0.0042) & (0.0024) & (0.1384) & (0.0210) & (0.0081) & (0.1609) & (0.0185) & (0.0099) \\
\hline\noalign{\smallskip}
\end{tabular}
\end{table} 

Based on the simulation results we may conclude that when there are redundant or noise variables a suitable subset of GMMDR variables allows us to significantly improve cluster identification. The improvement is larger for small samples having different within-cluster covariances and in the presence of noise variables. Overall, using GMMDR variables do not produce worse results than using original variables, but often leads to a better selection of the clustering structure. There appears no reason to perform model-based clustering on principal components. 

\subsubsection{Unequal prior probabilities}

The previous simulation examples assumed equal prior group probabilities. It may be of interest to investigate if the results obtained depend on such assumption. We expect that, provided there is enough cluster information in the data, no dramatic change from previous findings should be observed.
Thus, we replicated the previous simulation study for the more complex Model 3 (VVV), but with unequal prior probabilities set at $\pi = (0.1, 0.3, 0.6)$. 
The results are shown in Table~\ref{tab2:sim}.

When no noise variables are included the accuracy of GMM and GMMDR are essentially the same as in the equal prior probabilities case, whereas PCA+GMM performs slightly worse. If irrelevant variables (either noise or redundant ones) are present both GMM and GMMDR yield somewhat worse results when $n=300$ or $1000$. This seems to be due to the fact that BIC often selected models with more components than the true number of clusters.
On the contrary, PCA+GMM seems to improve when redundant variables (i.e. variables correlated with clustering ones) are included. Since principal components are computed from the correlation matrix, we argued that correlated variables are likely to provide useful clustering information when some groups are represented by few observations.
Overall, we note that GMMDR never performed worst than GMM, and it was also better than PCA+GMM except when irrelevant variables are included and $n=300$. 

\begin{table}[htb]
\centering
\caption{Average adjusted Rand index (with standard errors within parenthesis) based on 1000 simulations with unequal prior probabilities.}
\label{tab2:sim}
\smallskip\small
\setlength{\tabcolsep}{3pt}
%\renewcommand\arraystretch{0.8}
%\renewcommand{\tabcolsep}{1pt}
%\begin{tabular}{lp{0.2\linewidth}p{0.2\linewidth}p{0.35\linewidth}}
\begin{tabular}{l@{\hspace{10pt}}ccc@{\hspace{10pt}}ccc@{\hspace{10pt}}ccc}
\hline\noalign{\smallskip}
\textit{Model 3 (VVV)}  & \multicolumn{3}{c}{\textit{No noise}} & \multicolumn{3}{c}{\textit{Noise variables}} & \multicolumn{3}{c}{\textit{Noise and redundant variables}} \\
\hline\noalign{\smallskip}
& $n = 100$ & $n = 300$ & $n = 1000$ & $n = 100$ & $n = 300$ & $n = 1000$ & $n = 100$ & $n = 300$ & $n = 1000$ \\
\hline\noalign{\smallskip}
GMM 		& 0.9826 & 0.9991 & 0.9994 & 0.7451 & 0.6195 & 0.7407 & 0.8214 & 0.7326 & 0.7209 \\ 
    		& (0.0502) & (0.0025) & (0.0010) & (0.1468) & (0.1663) & (0.1160) & (0.1433) & (0.1188) & (0.0937) \\ 
PCA+GMM 	& 0.8781 & 0.8644 & 0.8606 & 0.8305 & 0.8347 & 0.7398 & 0.8334 & 0.8434 & 0.8228 \\ 
 		& (0.0970) & (0.0871) & (0.0532) & (0.0748) & (0.1050) & (0.2017) & (0.0590) & (0.0353) & (0.1054) \\ 
GMMDR 	& 0.9742 & 0.9991 & 0.9994 & 0.8733 & 0.8176 & 0.7964 & 0.8720 & 0.8073 & 0.8136 \\ 
 		& (0.0734) & (0.0024) & (0.0011) & (0.1341) & (0.1378) & (0.1547) & (0.1303) & (0.1601) & (0.1600) \\ 
\hline\noalign{\smallskip}
\end{tabular}
\end{table} 

\subsubsection{High-dimensional data}

Clustering high-dimensional data is a challenging task. The main problem in fitting  a GMM model is the large number of parameters which need to be estimated. 
To overcome this \cite{Ghosh:Chinnaiyan:2002} suggested to perform model-based clustering on the first few principal components. Based on the conclusion of \citet{Chang:1983}, discussed also in Section~\ref{drmbc}, this approach does not seem to be advisable. \cite{Bouveyron:Girard:Schmid:2007} proposed a method for dealing with high dimensional data.

In Section~\ref{simul2} we described a three clusters VEV model (2) on 3 clustering variables. We also considered the inclusion of 3 redundant variables (i.e., correlated with the clustering ones) and 4 noise variables; there was a total of ten variables according to the scheme $\{3|3|4\}$.
This basic setup has been generalized to investigated the behavior of the proposed approach as the number of variables increase. For $k=\{1,2,3,5\}$ we generated $10k$ variables following the scheme $\{3k|3k|4k\}$; for instance, if $k=3$, there are 30 variables, 9 of them are clustering variables, 9 are correlated with the previous ones, and the remaining 12 are noise variables.
We only investigated the most difficult case, i.e. the case of a small sample size, by setting $n=100$.

As the number of variables increase the accuracy of all the three methods tend to decrease, with GMMDR always more accurate on the average that the other two methods (see Table~\ref{tab3:sim}). 
However, for $p=40$ and $p=50$ the average ARI for GMMDR dropped just above $0.6$. Looking in more detail the results, we realized that, even though a subset of directions could provide better cluster accuracy, BIC frequently selected too many of such directions. To find evidence of this fact we replicated the simulation study, but using the simple model EII (common spherical covariance matrix) and selecting the GMMDR directions on the basis of the entropy criterion \citep{Celeux:Soromen:1996}. This provides a measure of the overlap of the clusters, and it is defined as 
$-\sum_{g=1}^{G}\sum_{i=1}^{n}t_{ig}\log(t_{ig})$, where $t_{ig}$ denotes the conditional probability that the $i$-th unit arises from the $g$-th mixture component.
Except for the case $p=10$, where the ARI is smaller, and $p=20$, where the ARI is essentially equivalent, the accuracy of GMMDR is largely increased (see Table~\ref{tab3:sim}). Note also that, although we constrained to fit the wrong EII model, the accuracy of GMM do not worsen.

Based on this limited study, it seems that for high-dimen\-sional data GMMDR directions are still able to recover the clustering structure of the data, but the BIC criterion overestimates how many of them are needed. When the entropy criterion is used to select relevant directions the clustering accuracy improve. 

\begin{table}[htb]
\centering
\caption{Average adjusted Rand index (with standard errors within parenthesis) based on 500 simulations for increasing dimensionality of the feature space and $n=100$. The last two rows show the results from fitting a GMM with a common spherical and equal volume model (EII) and the corresponding GMMDR with relevant directions selected by the entropy criterion.}
\label{tab3:sim}
\smallskip
\renewcommand\arraystretch{0.8}
\renewcommand{\tabcolsep}{1ex}
\begin{tabular}{l@{\hspace{10pt}}ccccc}
\hline\noalign{\smallskip}
\textit{Model 2 (VEV)}  & $p = 10$ & $p = 20$ & $p = 30$ &  $p = 40$ & $p = 50$ \\ 
\hline\noalign{\smallskip}
GMM & 0.6788 & 0.6405 & 0.6128 & 0.5968 & 0.5859 \\ 
 & (0.0850) & (0.0702) & (0.0744) & (0.0726) & (0.0783) \\ 
PCA+GMM & 0.4602 & 0.7238 & 0.6928 & 0.6485 & 0.6009 \\ 
 & (0.1831) & (0.1655) & (0.1574) & (0.1582) & (0.1283) \\ 
GMMDR & 0.9205 & 0.9717 & 0.8447 & 0.6802 & 0.6131 \\ 
 & (0.0843) & (0.0951) & (0.1960) & (0.2092) & (0.1920) \\ 
\hline\noalign{\smallskip}
GMM (EII) & 0.7101 & 0.6581 & 0.6299 & 0.6148 & 0.5979 \\ 
 & (0.0889) & (0.0707) & (0.0696) & (0.0753) & (0.0765) \\ 
GMMDR & 0.8447 & 0.9830 & 0.9817 & 0.9204 & 0.8381 \\ 
 & (0.1672) & (0.0837) & (0.0801) & (0.1528) & (0.1923) \\ 
\hline\noalign{\smallskip}
\end{tabular}
\end{table} 

An extreme form of high-dimensional data is the $p \gg n$ case. Clustering of gene expression data arising from microarray experiments is a typical application and model-based approaches have been proposed \citep{Yeung:Raftery:etal:2001}. In order to apply the approach discussed in this paper, a form of regularization is required for the inversion of the covariance matrix in \eqref{kernel}, such as one of those discussed in \citet{Bernard:Gardes:Girard:2009}.
However, these aspects require further studies.

\section{Data analysis examples}
\label{examples}

\subsection{Italian wines}
\label{wines}

\citet{Forina:1986} reported data on 178 wines grown in the same region in Italy but derived from three different cultivars (Barolo, Grignolino, Barbera). For each wine 13 measurements of chemical and physical properties were made, such as the level of alcohol, the level of magnesium, the color intensity, etc. The data set is avalaible at UCI Machine learning data repository \url{http://archive.ics.uci.edu/ml/datasets/Wine}. The following analyses were performed on a standardized scale.

Modelling all the available variables with a GMM selects the model VEI with eight clusters, which gives a small adjusted Rand index (see Table~\ref{tab1:wines}). This can be largely improved by the variable selection method proposed by \citet{Raftery:Dean:2006}. A comparable accuracy is also achieved by \citet{McNicholas:Murphy:2008}, who applied a class of parsimonious {G}aussian mixture models (PGMM) based on mixtures of factor analyzers; their selected model has $q=2$ latent factors and it is denoted as CUU \citep[see][Table~1]{McNicholas:Murphy:2008}. We note, however, that, despite the improvement on accuracy, PGMM selects a wrong number of clusters (see Table~\ref{tab1:wines}).

\begin{table}[htb]
\centering
\caption{Model-based clustering results for some models fitted to the wine data.}
\label{tab1:wines}
\begin{tabular}{lrlrr}
\noalign{\smallskip}\hline\noalign{\smallskip}
Method & Number of & Model & $G$ & ARI \\
 & features &  & & \\
\noalign{\smallskip}\hline\noalign{\smallskip}
GMM & 13 & VEI & 8 & 0.48 \\
GMM (best subset$^\dagger$) & 5 & VEV & 3 & 0.78 \\
PGMM  & 2 & CUU & 4 & 0.79 \\ 
GMMDR & 5 & VEV & 3 & 0.85 \\
\noalign{\smallskip}\hline
\end{tabular}\par
$^\dagger${\small\texttt{(Malic, Proline, Flavanoids, Intensity, OD280)}}
\end{table}

We applied the GMMDR approach to this data set and we obtained a final VEV model with three clusters estimated on five GMMDR directions. 
This model yields an adjusted Rand index equal to 0.85 (see Table~\ref{tab1:wines}) and the resulting confusion matrix is reported in Table~\ref{tab2:wines}.

\begin{table}[htb]%\sidecaption
\centering
\begin{tabular}{lrrr}
\noalign{\smallskip}\hline\noalign{\smallskip}
 & \multicolumn{3}{c}{Cluster} \\
Wines &  1 &  2 &  3 \\
\noalign{\smallskip}\hline\noalign{\smallskip}
Barolo     & 57 &  2 &  0 \\
Grignolino &  2 & 64 &  5 \\
Barbera    &  0 &  0 & 48 \\
\noalign{\smallskip}\hline\noalign{\smallskip}
\end{tabular}
\caption{A classification table for the best model found using GMMDR on the wine data.}
\label{tab2:wines}
\end{table}

\begin{figure}[htb]
%\sidecaption
\centering
\includegraphics[width=0.4\linewidth]{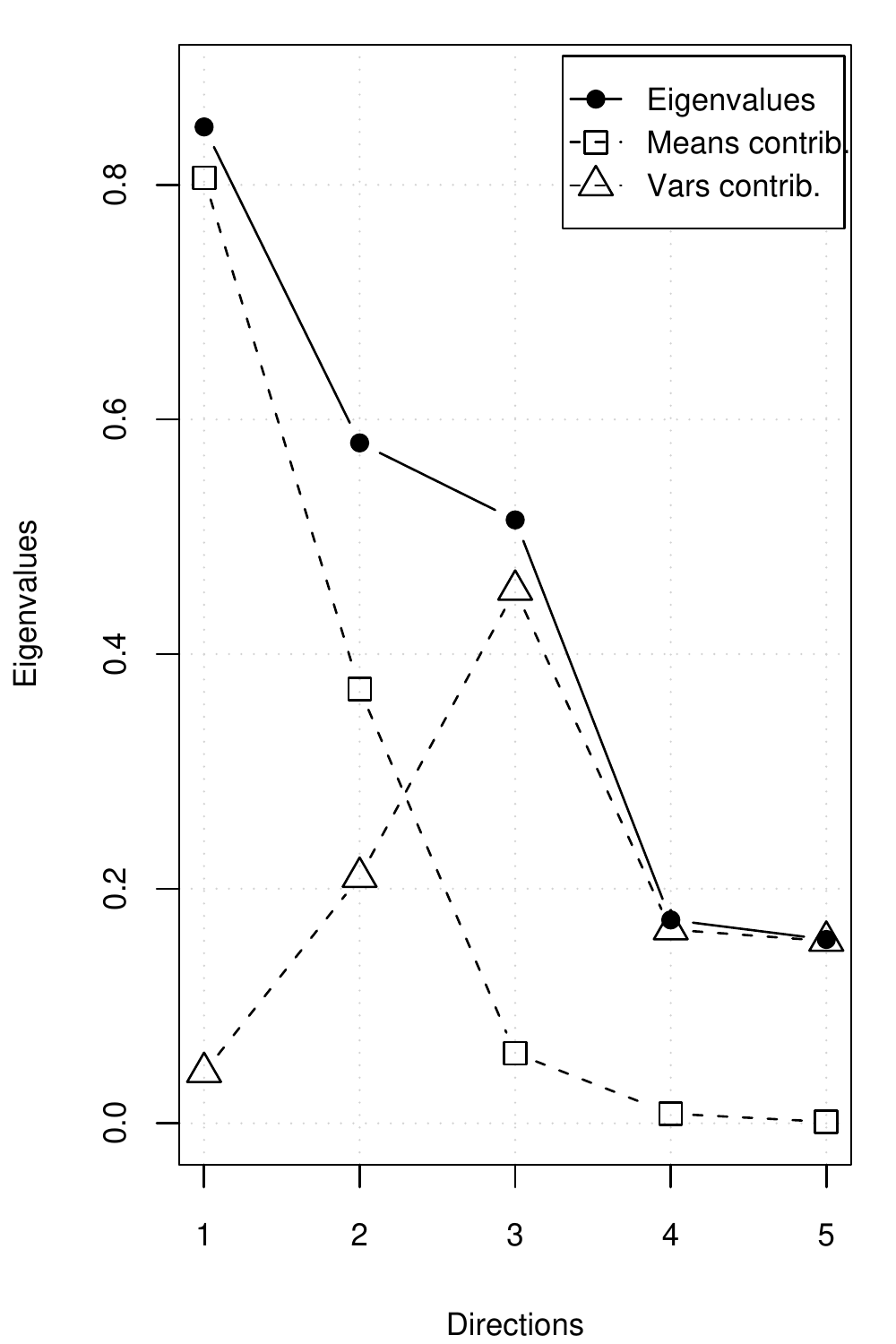}
\caption{Plot of eigenvalues for each GMMDR direction with contributions from means and variances differences among clusters for the wine data.}
\label{fig1:wines}
\end{figure}

As Figure~\ref{fig1:wines} shows, the first two directions mainly exhibit differences in cluster means, whereas the remaining directions mostly represent differences in cluster variances. We also note that the first three GMMDR directions appear to contain most of the clustering information.

\begin{figure}[htb]
\centering
\fbox{\includegraphics[width=0.8\linewidth]{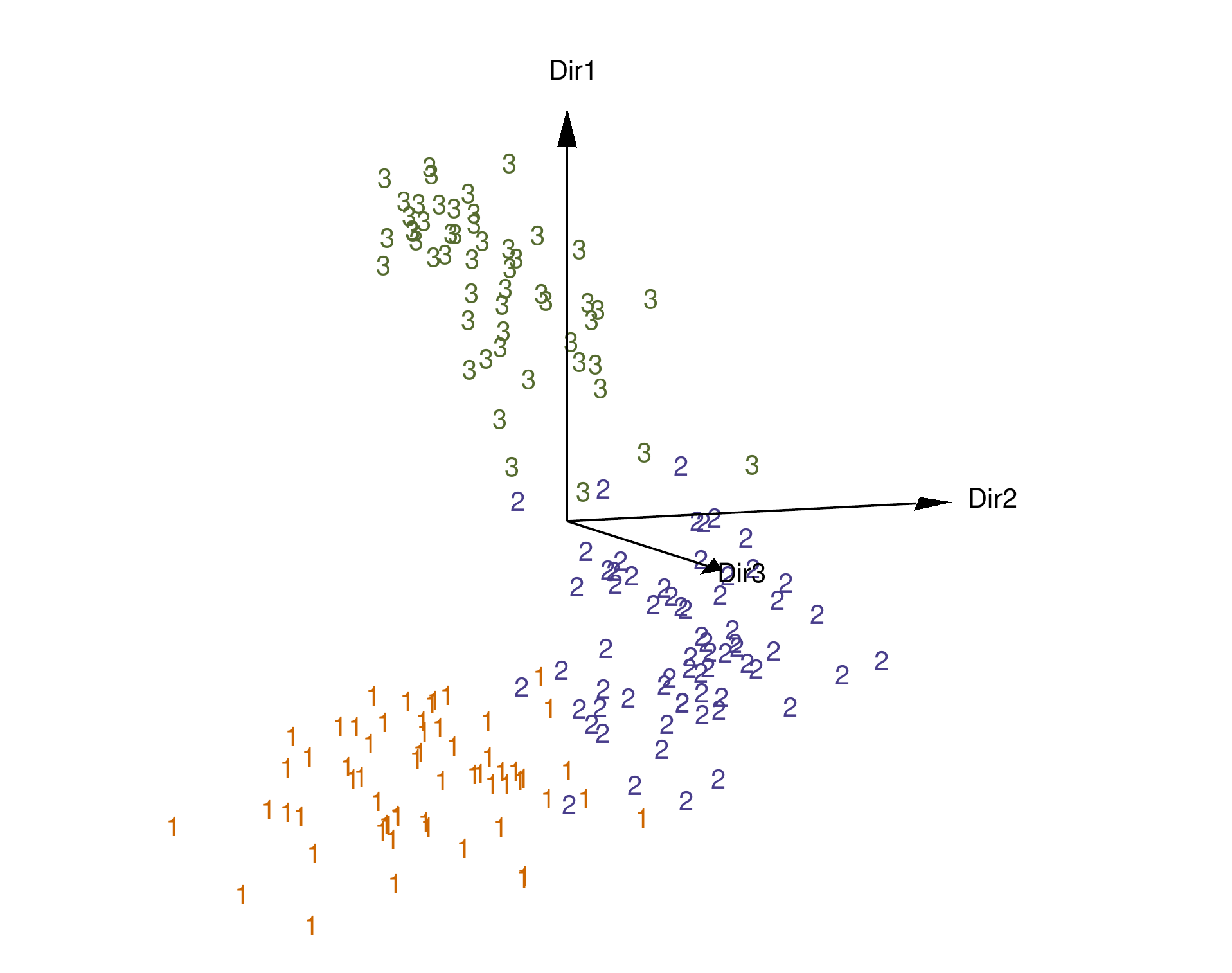}}
\caption{A static view of a 3D spinning plot of the first three GMMDR directions for the wine data. Data points are marked according to cluster membership; the clusters identified refer mainly to (1) Barolo, (2) Grignolino, and (3) Barbera.}
\label{fig2:wines}
\end{figure}

Figure~\ref{fig2:wines} shows a static view of a rotating 3D spin plot for the first three GMMDR directions with data points marked according to cluster membership; Barbera wines (3) appear to be well separated from the others along the first direction, while Barolo (1) and Grignolino (2) wines present a significant separation on the plane identified by the second and third directions.

\subsection{Australian crabs data}
\label{crabs}

This data set contains measurements recorded on 200 specimens of Leptograpsus variegatus crabs on the shore in Western Australia \citep{campbell:mahon:1974}.
Crabs are classified according to their colour form (blue and orange) and sex.
There are 50 specimens of each sex of each species. Each specimen has 5 measurements: frontal lobe size (FL), rear width (RW), carapace length (CL) and width (CW), body depth (BD). The main interest is to distinguish between species and sex based on the available morphological measurements.

Gaussian mixture modelling performs poorly for this da\-ta\-set. The model with the highest BIC selects 9 components, which gives a small adjusted Rand index. If we constrain the number of components to match the actual number of classes, the performance is still poor (see Table~\ref{tab1:crabs}).
Raftery and Dean (2006) obtained a large improvement through variable selection, ending up with a mixture model based on four variables. 
This dataset was also analyzed by \citet[pp. 516--517]{Bouveyron:Girard:Schmid:2007} who obtained a 5\% error rate with the high-dimensional data clustering (HDDC) model $[a_i b_i Q_i d_i]$.

Starting with the ``best'' unconstrained mixture model (EEE, $G=9$) we estimated the GMMDR directions and then we performed feature selection. The final GMMDR model is fitted on three directions with 4 clusters having ellipsoidal, equal volume and shape but different orientation (EEV). The error rate is 7.50\%, the same obtained by \citet{Raftery:Dean:2006}. The resulting accuracy is slightly worst than that for the HDDC model of \citet{Bouveyron:Girard:Schmid:2007}, but it has the advantage of providing a visualization of the clustering structure.

\begin{table}[htb]
\centering
\caption{Model-based clustering results for some models fitted to the crabs data.}
\label{tab1:crabs}
\renewcommand\arraystretch{0.8}
\begin{tabular}{llrrr}
\noalign{\smallskip}\hline\noalign{\smallskip}
Method & Model & $G$ & Error & ARI \\
 & & & rate (\%) & \\
\noalign{\smallskip}\hline\noalign{\smallskip}
GMM (unconstrained) & EEE & 9 &       & 0.4831 \\
% (unconstrained) & & & & & \\
GMM (constrained) & VEV & 4 & 42.50 & 0.3165 \\
%(constrained)  & & & & & \\
GMM (best subset$^\dagger$) & EEV & 4 & 7.50 & 0.8154 \\
HDDC & $[a_i b_i Q_i d_i]$ & 4 & 5.00 & \\
GMMDR  & EEV & 4 &  7.50 & 0.8195 \\
\noalign{\smallskip}\hline
\multicolumn{5}{l}{$^\dagger${\small\texttt{(CW, RW, FL, BD)}}}
\end{tabular}
\end{table}

\begin{figure}[htb]
\centering
\includegraphics[width=0.8\linewidth]{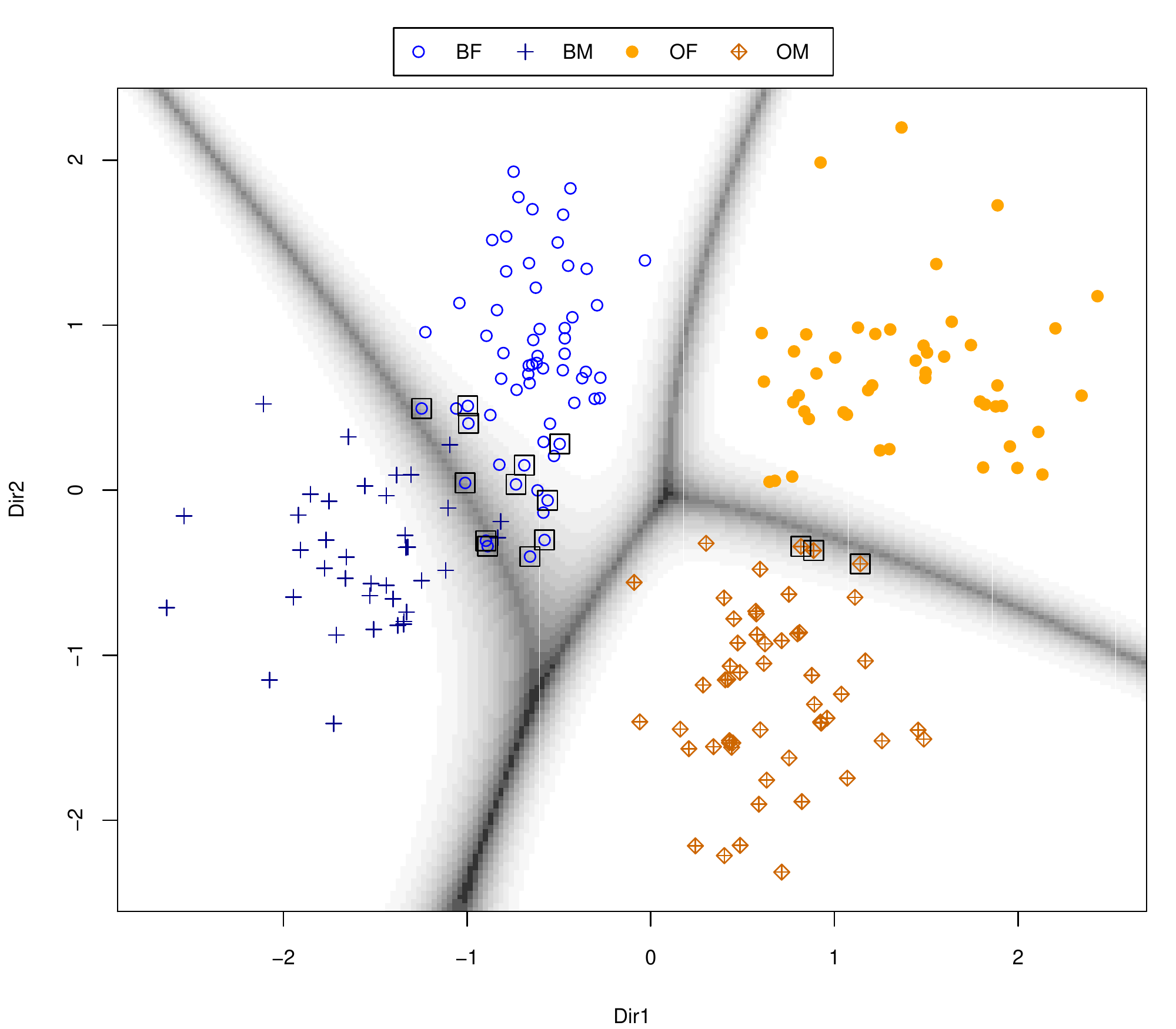}
\caption{Plot of crabs data points with uncertainty areas projected onto the first two GMMDR directions. Data points are marked according to the estimated cluster membership, while a square surrounding a point indicates a misclassified crab.}
\label{fig3:crabs}
\end{figure}

Summary plots (not shown here) indicate that the first direction mainly separates blue from orange type of crabs, while the second reflects sex differences. The last direction adds only marginal information contrasting blue and orange crabs within sex. Figure~\ref{fig3:crabs} plots the uncertainty $u_i = 1 - \max_g(\hat{z}_{ig})$ projected onto the first two GMMDR directions, where $\hat{z}_{ig}$ is the estimated conditional probability that observation $i$ belongs to group $g$.
Misclassified crabs are indicated by a square surrounding the point. As it can be seen, orange and blue crabs appear well separated, whereas discrimination of sex within crabs species is more problematic: some points lie in the regions of higher uncertainty (darker areas), in particular for blue crabs. It is interesting to note that these aspects can also be read-off from the corresponding confusion matrix (see Table~\ref{tab2:crabs}).

\begin{table}[htb]%\sidecaption
\centering
\begin{tabular}{lrrrr}
\noalign{\smallskip}\hline\noalign{\smallskip}
 & \multicolumn{4}{c}{Cluster} \\
Species &  1 &  2 &  3 &  4 \\
\noalign{\smallskip}\hline\noalign{\smallskip}
   BF & 50 &  0 &  0 &  0 \\ 
   BM & 12 & 38 &  0 &  0 \\
   OF &  0 &  0 & 47 &  3 \\
   OM &  0 &  0 &  0 & 50 \\
\noalign{\smallskip}\hline\noalign{\smallskip}
\end{tabular}
\caption{A classification table for the final GMMDR model fitted on the crabs data.}
\label{tab2:crabs}
\end{table}

%%%%%%%%%%%%%%%%%%%%%%%%%%%%%%%%%%%%%%%%%%%%%%%%%%%%%%%%%%%%%%%%%%%%%

%\clearpage
\section{Conclusion}
\label{conclusion}

In this paper we addressed the problem of dimension reduction for model-based clustering from the point of view of visualizing the clustering structure.
Often, principal component analysis is used as a pre-processing step, and a clustering procedure is applied to the leading principal components. Probabilistic principal components and factor analyzers have been also used for dimension reduction. However, these do not necessarily capture the underlying clustering structure. 

Unlike other approaches \citep{Tipping:Bishop:1999a, Tipping:Bishop:1999b, McLachlan:Peel:Bean:2003, Bouveyron:Girard:Schmid:2007, McNicholas:Murphy:2008}, the proposed method is not connected to any underlying latent variables structure, nor an EM-like algorithm is employed in the estimation. Instead, we aim at identifying the subspace which capture most of the clustering information contained in the data. Estimation of directions spanning such subspace is based on the eigendecomposition of a suitable kernel matrix, with unknown parameters estimated from a GMM selected as the best description of the data at hand. 
Visualization techniques can then be applied using the leading directions in order to obtain summary plots.
The estimated directions can be further reduced by recasting the problem of feature selection as a model choice problem and using BIC to approximate Bayes factors. A forward greedy search algorithm is discussed to avoid the search over all possible subsets.

The proposed methodology has been applied to simulated and real data sets. In all cases we were able to identify a dimension reduced subspace while preserving the cluster information contained in the original variables. Moreover, simulation studies showed that a suitable subset of directions allows for significant improvement in cluster identification. The improvement is larger when within-cluster covariances are different and redundant or noise variables are present.

Finally, it is straightforward to extend the proposed approach in a supervised context, where the class membership for each unit is known in advance.

%%%%%%%%%%%%%%%%%%%%%%%%%%%%%%%%%%%%%%%%%%%%%%%%%%%%%%%%%%%%%%%%%%%%%%%
\appendix
\section*{Appendix}
\renewcommand{\thesection}{A}
\setcounter{section}{1}  % reset counter 

\subsection{Proofs}
\label{appendix:proofs}

\paragraph{Proof of proposition \ref{prop:invariance}}

Define the transformed data matrix as $\X^* = \X\C + \ones\a\T$, where $\C$ is a $(p \times p)$ nonsingular matrix, $\a$ is a $p$-vector of real values, and $\ones$ is a $n$-vector of ones.
It is easy to show that in the transformed scale $\Sigmab^* = \C\T\Sigmab\C$ and the kernel matrix is given by $\M^* = \C\T\M\C$. The generalized eigendecompostion in the transformed scale satisfy the equation $\M\C\V^* = \Sigmab\C\V^*\L^*$, so $\C\V^*$ must be equal to the matrix $\V$ of eigenvalues of the generalized eigendecomposition of $\M$ with respect to $\Sigmab$, and $\L^*$ is equal to the corresponding diagonal matrix of eigenvalues. Hence, $\V^* = \C\inv\V$, $\L^*=\L$, and the GMMDR directions in the transformed scale are given by $\betab^* = \C\inv \betab$. 
Furthermore, the corresponding variates $\Z^* = \X^*\betab^* = \X\betab + \ones\a\T\C\inv\betab$, so they are simply a shifted version of the GMMDR variates in the original scale.

\paragraph{Proof of proposition \ref{prop:eigenvalues}}

For any kernel matrix we may rewrite the eigendecomposition in \eqref{gmmdr:decomp} as $\M\V = \Sigmab\V\L$. Since by definition $\V\T \Sigmab \V = \I_d$, the diagonal matrix of eigenvalues can be expressed as $\L = \diag[l_i] = \V\T\M\V$.

We start by considering the kernel matrix $\M_I$, for which $\M_I = \Var(\Exp(\X|Y))$. Thus, 
\begin{equation*}
\L = \V\T\Var(\Exp(\X|Y))\V = % \Var(\Exp(\X|Y)\V) = 
\Var(\Exp(\Z|Y)) = \diag[\Var(\Exp(Z_i|Y)],
\end{equation*} 
where $\Z = \X\V$ is the projection onto the subspace spanned by the columns of $\V$. So, for a SIR kernel matrix each eigenvalue is equal to the variance of the between group means along the $i$-th direction.
Consider the kernel matrix in \eqref{kernel}, for which the diagonal matrix of eigenvalues can be written as
\begin{equation*}
\L = % \V\T (\M_I\Sigmab\inv\M_I + \M_{II}) \V = 
\V\T\M_I\Sigmab\inv\M_I\V + \V\T\M_{II}\V.
\end{equation*}  
We note that since $\V\T\Sigmab\V=\I_d$ we have $\Sigmab\inv=\V\V\T$. Thus, the first part of the above equation may be written as
\begin{equation*}
\V\T \M_I \Sigmab\inv \M_I \V = (\V\T\M_I\V)(\V\T\M_I\V) = \Var(\Exp(\Z|Y))^2.
\end{equation*} 
For the following part we have
$$
\V\T\M_{II}\V =
\V\T \left[ \sum_{g=1}^G \pi_g (\Sigmab_g - \Sigmabar) \Sigmab\inv (\Sigmab_g - \Sigmabar)\T \right] \V = 
\Exp(\Var(\Z|Y)^2).
$$
Therefore, the matrix of eigenvalues can expressed as 
$\L = \Var(\Exp(\Z|Y))^2 + \Exp(\Var(\Z|Y)^2)$, 
and proposition \ref{prop:eigenvalues} follows.

\subsection{Greedy search algorithm for feature selection}
\label{appendix:greedysearch}

In this section we provide a detailed description of the greedy search algorithm used to select the ``best'' subset of GMMDR variates. 

The BIC criterion \citep{Schwartz:78} is the basic building block for choosing the number of clusters and model parametrization \citep{Fraley:Raftery:1998}. 
In general, for a finite mixture model $\Model$ with $g$ components it is computed as
\begin{equation*}
BIC(\Model, g) = 2 \max l - \nu_{\Model, g} \log(n),
\end{equation*}
where $\max l$ is the maximized log-likelihood, $\nu_{\Model, g}$ is the number of parameters estimated, and $n$ is the number of observations.
The difference between BIC values for two models is approximately equal to twice the logarithm of the Bayes factor for the comparison of the two models \citep{Kass:Raftery:1995}.
\citet{Raftery:Dean:2006} investigated variable selection in model-based clustering, recasting the problem as a model selection procedure. We adapt their approach for GMMDR variable selection.
 
Let $\Set = \{1,2,\ldots,d\}$ be the set of GMMDR variables $Z_\Set$ computed for a mixture model following the method described in Section \ref{estimation}. 
The criterion used for model comparison is the BIC difference in equation \eqref{BICdif}. 
The steps of the greedy search algorithm are the following:

\begin{enumerate}

\item Select the first variable to be the one which maximizes the BIC difference criterion. Let $\Set_1 = \{i\}$ be the candidate set, and $\Set'_1 = \emptyset$ be the set of already included variables, which is of course empty at the beginning.
We select the variable $Z_{i_1}$ such that
\begin{align*}
i_1 & = \arg_{i \in \Set}\max\; \BICdiff(Z_{\Set_1}) \\
    & = \arg_{i \in \Set}\max\; \left(\BICclust(Z_{\Set_1}) - \BICreg(Z_i)\right).
\end{align*}
At the first step $\dim(\Set_1)=1$, thus maximization is taken over the clustering models $E$ and $V$ with $G=1,2,\ldots,\max(G)$, where $\max(G)$ is the maximum number of clusters to be considered for the data.
Then, define $\Set_1 = \{i_1\}$, set $j=2$ and go to the next step.

\item Select a variable to include, among those not already included, to be the one which maximizes the BIC difference criterion. Formally, let $\Set_j = \Set_{j-1}\cup\{i\}$ be the candidate set, and $\Set'_j = \Set_{j-1}$ be the set of already included variables. We choose the variable $Z_{i_j}$ to be included such that
\begin{equation*}
i_j = \arg_{i \in \Set\setminus\Set'_j}\max\; \BICdiff(Z_{\Set_j})
\end{equation*}
In this case the ``best'' model is identified with respect to both the number of mixture components up to $\max(G)$, and model parametrization.
Then, update the subset of currently included variables, i.e. $\Set_j = \Set'_j \cup \{i_j\}$.

\item Set $j = j+1$ and iterate the previous step until a stopping rule is met.
The algorithm naturally terminates when all variables are included, but it might be stopped earlier when, for example, the maximal BIC difference for the inclusion of a variable becomes negative.
\end{enumerate}

As mentioned, the proposed greedy search is a forward-type algorithm. Given the orthogonality of the GMMDR variables, a backward step is not required.

%%%%%%%%%%%%%%%%%%%%%%%%%%%%%%%%%%%%%%%%%%%%%%%%%%%%%%%%%%%%%%%%%%%%%%%

%\begin{acknowledgements}
%\end{acknowledgements}

%:References

%\bibliographystyle{apalike}
%\bibliography{$HOME/Stat/mybiblio} % name your BibTeX data base

\end{document}